%% file: AutoRSP.tex
\documentclass[sigconf,9pt]{acmart}
\usepackage{xspace}
\usepackage{algcompatible}
\usepackage{algorithm}
\usepackage{stfloats}
\usepackage{subcaption}
\usepackage{natbib}

\renewcommand\footnotetextcopyrightpermission[1]{}
\newcommand{\sssec}[1]{ {{\flushleft \textbf{#1}}}}
\newcommand{\name}{AutoRSP\xspace}

\title{\texttt{AutoRSP:} An AI Agent for Radar Sensing Pipeline Design}

\author{Zhi Zheng\qquad Xincheng Xie\qquad Avery Gump\qquad Sacha Jungerman\\ Hongyang Li\qquad Yin Li\qquad Mohit Gupta\qquad Akarsh Prabhakara}
\affiliation{\institution{University of Wisconsin-Madison}\city{}\country{}}

\begin{abstract}
Radar sensing demands a pipeline that is well-thought-out across hardware and software. We ask whether an AI agent can automatically design radar sensing pipelines at a fraction of the time, cost, and expertise that current practice needs. We present \name, an agent that embeds radar-system-level structure through sub-agents mapped to five stack layers, a code-based representation of every layer with explicitly modeled cross-layer dependencies. Feedback for the agentic search process comes from a pipeline of realistic RF simulation, on-sensor emulation, and execution on real edge devices, rather than from an LLM judge. Because such feedback can be slower than an LLM call, we present methods to efficiently optimize agentic loops. We evaluate \name\ on robotic imaging, automotive perception, and vital sign sensing. It produces executable pipelines with performance comparable to or better than human-expert pipelines and closed-source agents. \name's harness, tools, and benchmarks will be released to bridge the rising gap in agentic research abilities outside of frontier labs. \vspace*{-0.1cm}
\end{abstract}

\begin{document}
\pagestyle{plain}

\maketitle
{\renewcommand{\thefootnote}{}\footnotetext{Email: \{zzheng, xxc, hli2269, mohitg, akarsh\}@cs.wisc.edu, \{gump, sjungerman, yin.li\}@wisc.edu}}

\input{introduction}

\input{design}
\input{evaluation}

\input{conclusion}

\bibliographystyle{ACM-Reference-Format} 
\bibliography{references}

\end{document}

%% file: introduction.tex
\section{Introduction}
\label{sec:intro}

Compact radars are becoming indispensable in automotive and robotic sensing. Each new application demands a rethinking of the radar sensing pipeline, and the rethinking cuts across hardware and software layers. A logistics firm extending through-box perception to glassware adapts its processing algorithms to specular targets \cite{NLOS}. A driving-assistance team facing the same specular physics on car bodies instead distributes several radars across the bumper, so that a fade-out at one viewpoint is covered by another \cite{bansal2020pointillism}. The same physical problem admits two remedies at opposite ends of the stack. Choosing between them means co-designing hardware, firmware, software, and system integration under tight constraints on \underline{s}ize, \underline{w}eight, \underline{a}rea, \underline{p}ower, and \underline{c}ost (SWaP-C), all while meeting a use-specific performance target. Mature organizations manage this with structured design and test cycles that cost people, time, and money, and that presume radar expertise to begin with.

Three scenarios illustrate the bind: (1) A robotics startup wants to use radar, but no one on the team is familiar with radars. (2) A team building an indoor occupancy sensor holds three plausible fixes for specular fade-out: widen the aperture, exploit multipath, or integrate over ego-motion. Which one wins is set by the incidence-angle, roughness, and motion statistics of their scenes, and those statistics become clear only once a pipeline exists to measure them. However, they have the capacity to build only one. (3) A group already shipping a radar pipeline has both the expertise and the capacity, yet keeps reaching for the same family of techniques that worked before. The design that would beat their baseline may sit outside the range they are comfortable searching, and nothing in their process would surface it. This paper asks: \textit{``how can we democratize access to holistic system-level radar expert knowledge, rapid prototyping of ideas, and research beyond current practice, at a fraction of the time and cost that current practices demand?"}

\begin{figure}%
\centering
\includegraphics[width=0.9\columnwidth]{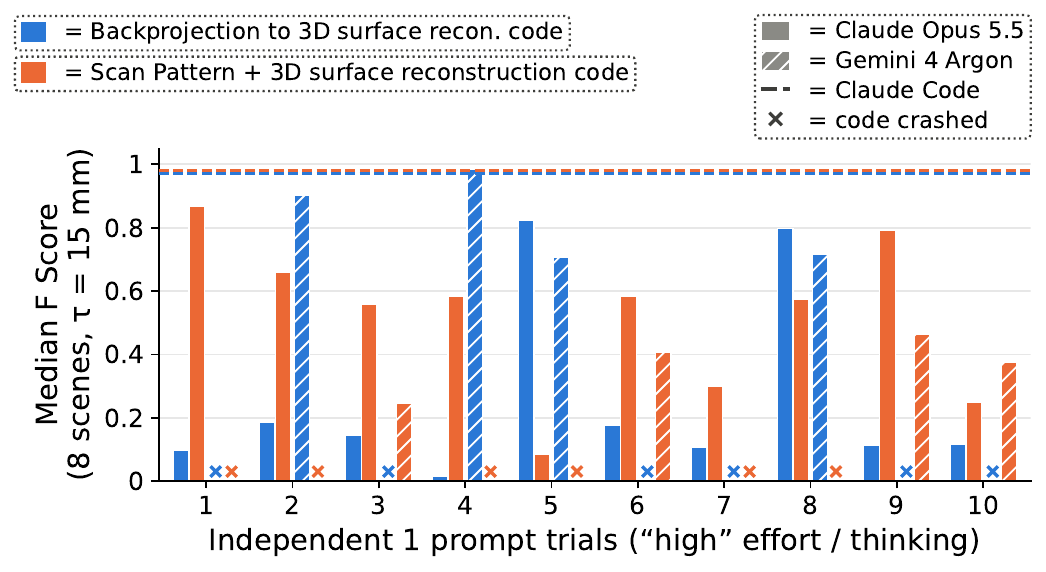}
\vspace*{-0.1in}
\caption{Motivating study shows that with repeated trials, single-prompt requests on a frontier model can yield high-quality solutions for a scoped radar design task, as indicated by higher top F-scores. However, the performance degrades on a more complex task that requires thinking across radar system layers. Frontier models in an agentic framework (Claude Code with Claude Opus 5.5 as model) can think across layers and return a very high score on a scoped task.}
\vspace*{-0.1in}
\label{fig:motivation}
\end{figure}

\begin{figure}%
\centering
\vspace*{-0.1in}
\includegraphics[width=0.9\columnwidth]{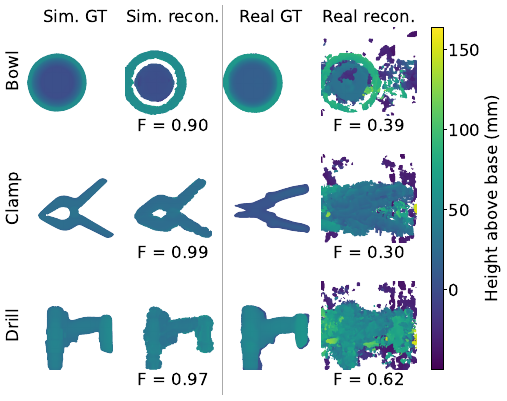}
\vspace*{-0.2in}
\caption{Even on a scoped one-layer question, we find that pipelines generated by frontier agents report good scores on high fidelity simulated measurements, but their solution fails on real objects. This occurs because the generated pipelines overfit to the success on agent's self-written simulations and evaluation tools, and fall prey to the sim-to-real gap.}
\vspace*{-0.25in}
\label{fig:motivation_v2}
\end{figure}

Large Language Models (LLMs) have succeeded at analogous design and discovery tasks, from code synthesis \cite{sweagent, autocoderover, swebench} to computer architecture search \cite{sankaralingam2026computer}, system optimization \cite{wang2024netconfeval,glia,karimi2026improving}, and Internet of Things design \cite{shen2025autoiot,ma2026autorf,zhou2025iot}. This success stems not only from stronger language models, but from inference-time infrastructure that steers and disciplines their output. Prompt and context engineering, multi-LLM orchestration, and ideate-evaluate-and-evolve loops have turned LLM-based design from a monolithic API call into a infrastructure-design-around-LLM problem. We ask what infrastructure around state-of-the-art LLMs (together, an "agent\footnote{A model could internally have its own agentic setup and most frontier models do today. We consider the infrastructure around models that we can control and build.}") is needed to bring this success to radar sensing pipeline design.

\begin{figure*}[t]
\centering
\includegraphics[width=\textwidth]{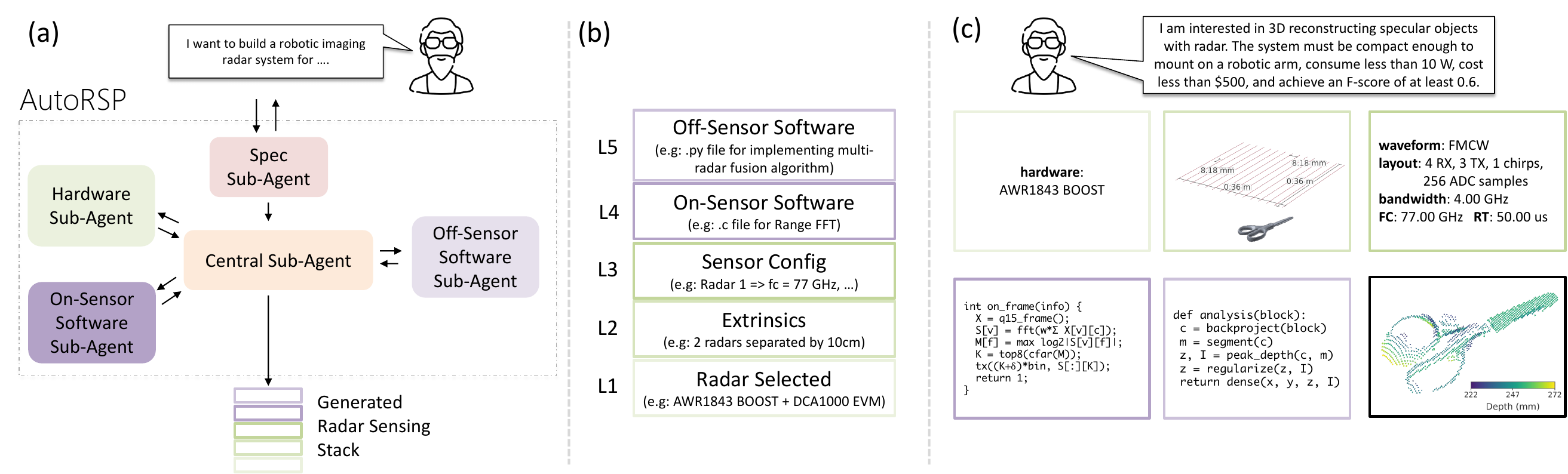}
\vspace*{-0.7cm}
\caption{(a) Interaction with \name\ to design a radar sensing pipeline. (b) The hardware-to-software stack for radar sensing. (c) \name's result for a robotic imaging design request. Each layer instantiated by \name\ is shown. 
}
\vspace*{-0.3cm}
\label{fig:hero}
\end{figure*}
We begin with a motivational study of how today's models and agents fare on radar sensing problems. Consider 3D surface reconstruction of specular targets, as in mmNorm~\cite{NLOS}. The design spans which radar to select, how to configure it, how to synthesize an aperture, and what code turns scans into a 3D surface, subject to requirements on time-to-acquisition, time-to-result, SWaP-C, and F-score. We first fix every design choice but one, keep only the F-score requirement, and prompt for the code that maps the backprojection image to a 3D surface. Over repeated single-prompt trials, a frontier model (Gemini 4 Argon / Claude Opus 5.5 at ``high'' effort) reaches an F-score as high as 0.98 (Fig ~\ref{fig:motivation}). Opening a second, coupled question (scan pattern alongside processing code) lowers the best score to 0.87. Their performance also varies widely across trials. Agents such as Claude Code~\cite{anthropic_claude_code}\footnote{Version 2.1.304 with Claude Opus 5.5 at ``high'' effort.}, which add context engineering, multiple model calls, and tool use, reason across layers, and reach 0.97 on the one-question task when evaluated on our high-fidelity simulator. Three problems remain. The agent does not natively know the structure of a fully executable radar pipeline, its solutions fail on real-world data because they overfit to the simple simulations it writes for itself (Fig~\ref{fig:motivation_v2}), and its judgement of compute-within-constraints is poor (shown in Sec~\ref{sec:eval}). Moreover, such agents are closed-source, which leaves researchers outside of frontier labs little capability to do agent-based research. Open-source design agents~\cite{openevolve} lag behind, as our evaluation shows (Sec~\ref{sec:eval}). Capable as they are, today's models and agents are not sufficient for system-level radar design.

We present \textbf{\name}, an AI agent for designing radar sensing pipelines. \name encodes human insight in the agent's structure (its ``harness'') and grounds it in reality through a tool suite of realistic simulation, emulation, and real-world execution. The interaction mirrors other design agents (Fig~\ref{fig:hero}a). The user specifies the problem, the requirements, and the edge device attached to the radar(s). 

\name\ makes key design decisions along three directions:

\sssec{(1) Embedding system-level structure.} The agent should ideally represent the entire pipeline, from hardware to software, and jointly search for superior designs across the stack. We first propose to breakdown the design problem to manageable chunks, then represent solutions at each stack layer in an LLM-friendly form, and embed the dependency across layers.

\underline{\textit{Sub-agents for layered breakdown:}} We layout five key layers that are of most interest to system designers: off-the-shelf radar(s) selection (L1), extrinsics (L2), sensor configurations (L3), on-sensor processing (L4) and off-sensor processing at an edge device (L5), see Fig~\ref{fig:hero}b. We leverage multi-LLM orchestration to split tasks between a central sub-agent that acts as a "hub", and "spokes" -- a hardware sub-agent, an on-sensor software sub-agent, and an off-sensor software sub-agent. Sub-agents\footnote{We refer to these as sub-agents as opposed to multi-agents because outside of our framework they may not be useable as is.} are not merely a simple LLM API call, but a careful composition of best-effort ideas: from prompting and context building, to efficient validation by interaction with tools for physically grounded context, or secondary LLMs as reviewers. Although this breakdown is by no means optimal or novel, it enables meaningful output generations, thus providing a platform to engineer infrastructure around. 

\underline{\textit{Code based layer output representation:}} Software sub-agents can design L4 and L5 as an executable code. But, a key challenge is in representing the other non-software layers. For example, a Synthetic Aperture Radar (SAR) imaging task would need L2 to output all the positions that the radar should be moved over. At a Nyquist sampling limit, this can be tens of thousands of extrinsics. This would blow up the output token size of the LLM responsible for generating L2. Our insight is to pose design problems at each stack layer as a coding problem, since LLMs excel at this, for compactness. 

\underline{\textit{Cross-layer structured dependencies:}} Choices across the stack have structured dependencies. For example, selecting a mmWave radar device requires L5 sub-agent to consider its wavelength while writing the code, or that selecting a uniform sampling grid for SAR could unlock faster L5 processing (frequency domain backprojection, instead of time). Thus, when the agent decides to update any one layer, it is important to explicitly control the relevant historical context based on currently frozen bottom layers. This disciplines the output from being mutually incompatible across layers.

\sssec{(2) Physically grounded evaluation.} A pure coding design agent, e.g., one solving LeetCode problems, focuses on producing a working solution and then improving. A cyber-physical pipeline should obtain a feedback score by executing the stack in physically meaningful scenes and on hardware. Clearly, this is time-consuming. A common workaround is to ask LLMs to serve as judges, decreasing the full-stack emulation time to a few LLM API calls. We instead resort to building a pipeline of physically realistic RF simulator, on-sensor emulation and off-sensor dispatch on a rig of real edge devices. Generated sensor configurations and extrinsics are also checked if they are physically realizable. This high quality feedback distinguishes our search from other general purpose agent. Worth noting is our approach to full-stack emulation. While one may be tempted to do compute emulation via all-in-simulation or all-on-real-devices, we take a middle ground: simulation for on-sensor compute that have a few standard popularly used programmable routines, and real-check for off-sensor compute where a wide variety of software approaches can be used on radar-data. 

\begin{figure}
\centering
\includegraphics[width=0.9\columnwidth]{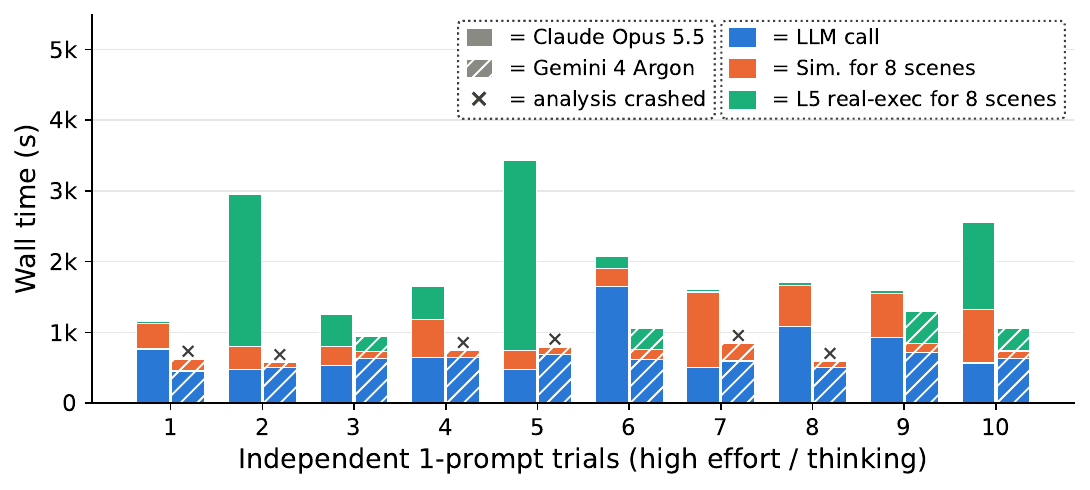}
\vspace*{-0.1in}
\caption{
RF simulation and processing code execution (on NVIDIA 5090) for grounded evaluation are not cheap. The queries here are for the same 2-question task seen in Fig~\ref{fig:motivation}. Evaluation time can be quite substantial compared to an LLM query in some cases. The LLMs did not output dense scan patterns here. Denser scan patterns when suggested by the LLM can take even longer to simulate and execute.
}
\vspace*{-0.25in}
\label{fig:time}
\end{figure}

\sssec{(3) Efficient agent interaction.} 
Our goal is to encourage the agent to explore many promising ideas under a fixed compute and time budget.
Thus, we consider efficiency from two standpoints.

\underline{\textit{Efficiency during evaluation:}} Tool usage for emulation and evaluation starts becoming the key bottleneck if every generation needs to get accurate feedback. Fig~\ref{fig:time} shows the time spent by an open-loop single-prompt query and contrasts it with the amount of time it would have taken to simulate radar measurements and execute the processing code to get feedback for next generation in a closed-loop agent. We observe that there is plenty of opportunity for efficiently using given resources to pipeline LLM calls, emulation and evaluation across layers. We support two modes which the user can configure at the beginning -- a pipelined architecture that doesn't compromise on accurate feedback, and one that trades-off accurate feedback (on recent generations) for generating more ideas via LLM calls and evaluating with LLM-as-judges. Next, inspired by human teams that progressively scale the difficulty of scenes-of-interest, rather than evaluating on all scenes in every generation, we use progressive-evaluation as planned by the central sub-agent or pre-defined by the system user. 

\underline{\textit{Efficiency due to agent structure:}} There exist three optimization opportunities. (1) Tiny modifications to off-sensor sub-agent output, like threshold parameters -- can invoke a complete re-run at the agent level. Radar processing algorithms are dominated by such parameters, for accurate detection and estimation based on scene properties and sensor noise, and prone to this. (2) Sub-agent code that is unoptimized to the edge is a waste of time during evaluation. (3) Re-simulating measurements from the same extrinsic or re-running a software block on same measurements across generations is unnecessary. We integrate an LLM-based parameter sweeper, an explicit off-sensor code optimizer, and cross-layer, cross-generation cache management to address these inefficiencies. 

\sssec{Evaluation.} We evaluate \name\ on tasks in robotic imaging (see one of the results at Fig~\ref{fig:hero}c), automotive perception, and vital sensing. We choose these tasks to demonstrate that our agent can innovate at the software layer and lower down at the hardware choice, or sensor configuration or placement. We give a number of creative design questions as prompts and show that \name\ successfully generates executable human-expert and better-than-human pipelines. We compare against strong baselines (Claude Code and OpenEvolve). Particularly, we also show hybrid approaches where important LLMs in \name\ are turned into individual coding-agent (Claude Code) calls, and when our custom written tools and insights are handed over to today's Claude Code. Our evaluation reveals that we can achieve performance comparable to or better than human-experts and closed-source agents.

\sssec{Scope.} Given the large scope, as a first step, we make the following choices to limit the scope. (1) We restrict our agent to pick from a list of COTS hardware. Designing all the way from ICs, PCBs and adding metasurfaces on radars is out-of-scope here. (2) Cutting-edge software sensing pipelines often tend to be learning-based today. We restrict our agent to push the frontier with physics-based signal processing pipelines, which is still very valuable, as it continues to be prominent in research and products. (3) We focus on radar(s)-only instead of multimodal fusion. The same layered structure is also faced by other sensing modalities and our approach can shed light there. (4) Non-radar parallel workflows can run on the edge or off-load from the edge to the cloud. We plan on expanding the scope to cover all these in the future. 

\sssec{Contributions.} \name\ is the first radar sensing pipeline design agent. Since we formulate everything radar-related as code, one could posit that harness design, high-quality tool creation, and tool access are also code. This raises the question: ``can everything be automated and subsumed under a future Claude-Code-like coding agent or even a future model?'' Given how rapidly agent design is evolving, this is plausible, but our work remains useful today and more broadly through: (1) formulations and code-based abstractions for a successfully executable radar pipeline, (2) evidence of which tools matter for successfully realizing a radar task, (3) a reference for a useful radar design agent today, and (4) a radar system design benchmark (prompts, scenes, evaluating metrics and scripts) for future model releases to compare against.

%% file: design.tex
\section{Background and Related Work}
This section provides a background on AI agents and related work in wireless and radar design.

\sssec{LLMs.}
LLM capabilities stem from the transformer~\cite{transformer}, instruction fine-tuning~\cite{flan, instructgpt}, and human feedback~\cite{rlhf}. At inference, chain-of-thought elicits intermediate reasoning~\cite{cot}, decomposition splits a hard goal into ordered sub-problems~\cite{leasttomost}, majority-vote sampling improves consistency~\cite{selfconsistency, verifier, letsverify}, and self-reflection lets a model critique and revise its output via feedback~\cite{selfrefine, reflexion}. Reasoning-trained models already incorporate these strategies~\cite{o1, deepseekr1}.

\sssec{From LLMs to agents.}
A problem can be routed through several role-playing LLMs instead of one monolithic call~\cite{camel, metagpt, llmjudge}, and their debate or cross-examination improves factuality and robustness over any single response~\cite{debate}. LLMs can also \emph{act}, interleaving reasoning with tool calls, code execution, and environment feedback~\cite{react, toolformer}. Frameworks orchestrating such reasoning-acting loops and multi-agent collaborations~\cite{autogen} make it practical to cascade pipelines for long-horizon tasks no single forward pass could solve. FunSearch~\cite{funsearch}, AlphaEvolve~\cite{alphaevolve,openevolve} and GEPA~\cite{agrawal2026gepa} couple an LLM with an automated evaluator in an evolutionary loop that repeatedly proposes, runs, and scores candidates, discovering algorithms that surpass human baselines and establishing that an LLM can search a program space against a measurable objective. Together, these ideas define the agentic paradigm~\cite{sweagent, autocoderover, swebench}.

\sssec{Systems for agents.} 
\cite{mcp,mcp_hw} streamline LLM--tool interface definitions.
\cite{tian2026agentprog,karimi2026improving} address long-horizon issues of design agents. \cite{ling2026timelyllm} optimizes latency for time-critical physical agents (robots and HCI, not auto-design). \cite{zhang2026kvswap,chung2026agent} optimize LLM inference for agentic interactions on personal devices.

\sssec{Agents for RF design.}
RF hardware design agents couple LLMs to EDA simulators~\cite{menter, rfampdesigner,ma2026autorf}. At the electromagnetic layer, \cite{metaagentic, mcpmetaoptics} couple LLMs to physics solvers for metasurface design. Recent LLM agents target IoT hardware generation~\cite{luo_iotgen}, network planning~\cite{radiomap}, IoT configuration and policy generation~\cite{collabiot}, and end-to-end wireless optimization with solvers in-the-loop~\cite{comagent}. \cite{ke2025signalllm} builds a signal processing design agent. Julia Dyad~\cite{dyad} integrates physics-based debugging and simulators with LLMs. \cite{behari2026task} optimizes sensor configurations in a 6D space to satisfy user requests, but ignores algorithms.

\sssec{Automated RF design prior to AI agents.}
Waveform design \cite{rlwaveform}, antenna search \cite{hornby2006automated}, array design \cite{wang2003optimal}, and DSP-transform platform-code generation \cite{puschel2005spiral} are well studied. Cognitive radar~\cite{haykin2006cognitive, bell2015cognitive}, adapting via end-to-end feedback, is closest in spirit to ours, but largely for on-the-fly adaptation, not de novo design.

\begin{figure*}[t]
\centering
\includegraphics[width=0.95\textwidth]{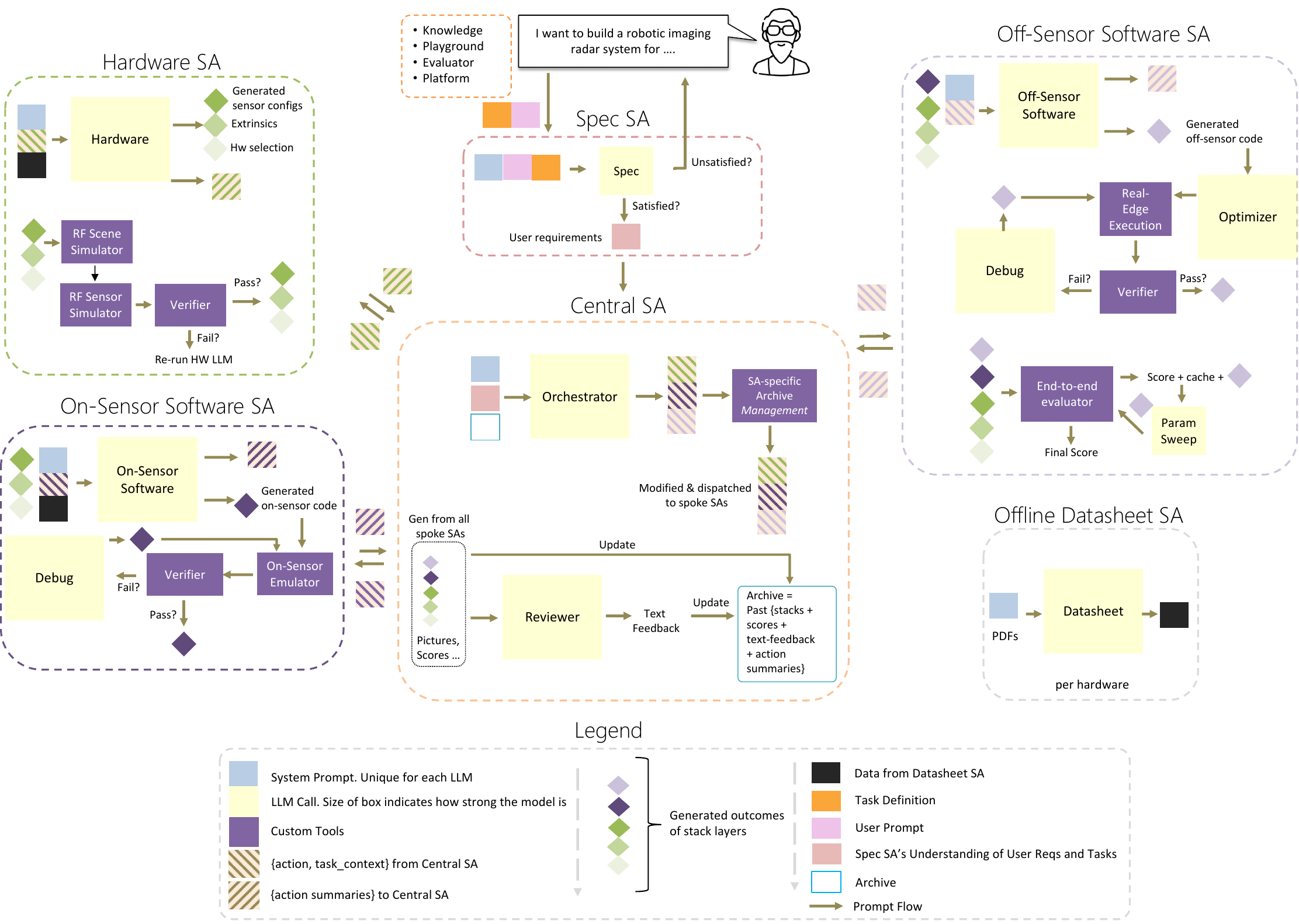}
\caption{This figure shows a detailed view of all the sub-agents (SA) managed by our agent, \name. Building an agent involves designing structured interactions between LLMs, code, tools, and prompts. The flow first goes from the user to the Spec SA. The Central SA then drives actions that go from hardware SA, on-sensor software SA to off-sensor software SA. An evaluator provides an overall score for the entire generated stack. An archive keeps memories of past scores and stacks. The orchestrator uses the archive and suggests a new plan every iteration.}
\vspace*{-0.3cm}
\label{fig:overview}
\end{figure*}

In wireless and radar, LLM-guided design takes a component-specific view of the system stack. We embrace the full stack, where choices at different layers can affect design outcomes. General-purpose, versatile agents like OpenEvolve and Claude Code (which have absorbed the best ideas from individual papers) are natural baselines that we compare \name\ against.

\section{\name\ Design}
\label{sec:design}

\subsection{Embedding System-Level Structure}
\label{ssec:system-structure}

\subsubsection{\underline{Sub-agents for layered breakdown}}
\name\ decomposes full-stack radar design hub-and-spoke: a Central Sub-Agent (SA) hub plans candidate designs, and ``spoke SAs,'' each owning a slice of expertise, execute its plans. Hardware SA owns L1--L3, and On-Sensor and Off-Sensor Software SAs own L4 and L5, generating code running on the radar sensor and user-chosen edge device.

\sssec{Anatomy of a sub-agent.}
Each SA has a unique system prompt, one or more LLM API calls, and custom tools for (1) RF world and sensor simulation, on-sensor emulation and dispatch for off-sensor edge execution; (2) verifiers that check whether a generated artifact is realizable or executable; and (3) end-to-end evaluation.

\sssec{Interactions.}
Fig~\ref{fig:overview} shows our harness (structure and control flow).

\textbullet\hspace*{0.1cm} A user request first enters the user-facing Spec SA, which elicits missing SWaP-C and performance details, and outputs a structured specification, looping with the user until the request is well-posed. Optional inputs are a reference subject-matter-expert knowledge base (\textit{Knowledge}), design-time scenes (\textit{Playground}), evaluator functions (\textit{Evaluator}), and misc.\ platform details like robot arm or car type, or configured edge-device power mode (\textit{Platform}).

\textbullet\hspace*{0.1cm} Central SA's \emph{Orchestrator} LLM reads the specifications and the \emph{Archive} (record of past agent-generated candidates) and proposes a plan. After a \emph{Management} tool handles cross-layer dependencies, it dispatches an \texttt{\{action, task\_context\}} message, specifying what each spoke SA should focus on, to the Hardware, On-Sensor and Off-Sensor Software SAs in order.

\textbullet\hspace*{0.1cm} Each spoke SA uses relevant earlier SAs' outputs (e.g., on-sensor software relies on hardware) during generation and local verification. Some SAs draw facts from the Datasheet SA, which summarizes each datasheet offline as datasheets are static and in several cases very long (100s of pages), needing longer LLM inference. Software SAs' \emph{Debug} LLMs iterate until outputs pass checks.

\textbullet\hspace*{0.1cm} An end-to-end evaluator runs the generated stack (RF and sensor simulation $\rightarrow$ on-sensor emulation $\rightarrow$ real edge execution of off-sensor code) on scenes for feedback.

\textbullet\hspace*{0.1cm} Beyond quantitative scores for user-requested aspects, a multi-modal \emph{Reviewer} (in Central SA) analyzes all generated code, scores and qualitative pictures (point clouds, heatmaps, etc.) to generate per-layer and cross-layer text feedback for \emph{Orchestrator} (inspired by~\cite{yuksekgonul2024textgrad}). Scores, feedback, and the full stack return to \emph{Archive}.

\textbullet\hspace*{0.1cm} Our agent evolves designs by updating each LLM's input prompt (most prominently, \emph{Archive} for \emph{Orchestrator}), repeating until user specifications are met or for a fixed number of iterations. We refer to iterations sometimes as generations.

\subsubsection{\underline{Code based output representation}}

\emph{On-Sensor} and \emph{Off-Sensor Software} LLMs generate C and Python code respectively, i.e., code synthesis. The \emph{Hardware} LLM's output is yet to be defined.

A simple structure is to have the Hardware SA output, in JSON: (1) L1: radar hardware picked from the local datasheet repository, (2) L2: an array of radar positions and orientations at which to simulate measurements, (3) L3: sensor configurations the chosen hardware needs. JSON is structured enough to pass around during interactions, but radar tasks can have very long L2, and consequently L3, outputs. For e.g, for SAR imaging, the Hardware LLM must generate all extrinsics the radar moves over, plus an L3 configuration for each: possibly tens of thousands of positions. This quickly blows up output tokens, input tokens of downstream LLMs (\emph{On-Sensor} and \emph{Off-Sensor Software}), and the size of the archive storing all outputs.

\begin{figure}[!htb]
\centering
\vspace*{-0.2cm}
\includegraphics[width=0.9\columnwidth]{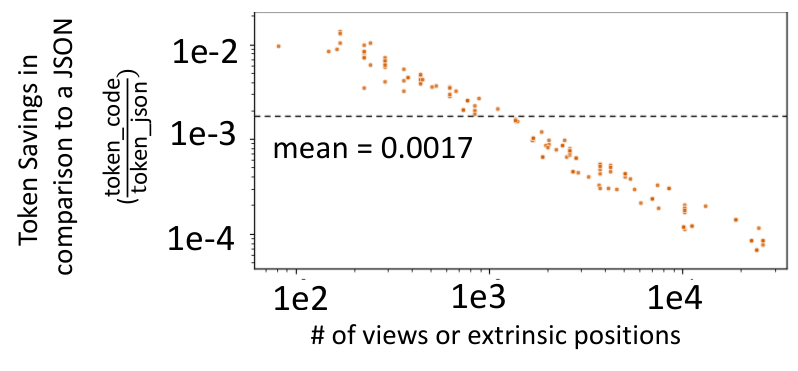}
\vspace*{-0.2in}
\caption{Efficacy of code based representation for L2}
\vspace*{-0.2in}
\label{fig:coding_problem}
\end{figure}

Instead, we find it best to pose this as a coding problem that generates the positions upon execution. For example, a uniform grid sampling strategy is a few lines of code with a spatial sampling parameter and grid size. Likewise, if the LLM varies sensor configurations (L3) spatially, that is a code block parameterized by spatial position. Enforced in the \emph{Hardware} LLM, this simple strategy consumes on average $600\times$ fewer tokens than JSON for L2 (Fig~\ref{fig:coding_problem}).

\subsubsection{\underline{Code-based scene generation}} 

We apply the same strategy to scene generation. Rather than hand-authoring every scene or relying only on existing ones from the literature, we give a coding agent reusable scene assets. It creates a shared library of parameterized functions and writes compact scene programs composing them with appropriate parameters, rather than enumerating geometry or motion frame by frame. This yields diverse scenes with manageable code and output size. For dynamic scenes, each scene XML must contain a Python code block defining a \texttt{frame(\allowbreak frame\_\allowbreak times)} generator that, for each requested timestamp, outputs a static scene describing the geometry at that instant for RF simulation, without listing every frame.

\subsubsection{\underline{Cross-layer structured dependencies}} A sub-selection of the most useful \emph{Archive} entries is typically input to \emph{Orchestrator}. Our selection policy explores or exploits at every iteration, and when exploiting, it selects the ``top-1 design for each design-time scene'' (other policies, like top-k, could be used). Under any policy, exploitation raises an important issue. If \emph{Orchestrator} updates only the software layers in the current generation, the entries selected from \emph{Archive} may have hardware different from the current generation's, and supplying this irrelevant information anyway may cause cross-layer incompatibility. The same holds when \emph{Orchestrator} freezes L1 to L4. With frozen bottom layers, we therefore further sub-select designs with a full bottom-layer match. \emph{Management} tool matches bottom-up for efficiency, searching over the $P$ design first selected by our policy ($P$ is the number of design-time-scenes).

\subsection{Physically Grounded Evaluation}
\label{ssec:physics-eval}

A dedicated pipeline of RF simulators, emulators and real executors provides physically grounded evaluation.

\sssec{RF Simulation.}
For every scene, radar extrinsic position and chosen hardware, we perform RF simulation (extract phasors) and sensor simulation (phasors to digital signals). Instead of extremely slow full-wave solvers (e.g., Ansys HFSS), we use ray-tracers with as many wave-like properties added, so simulation is fast and physically accurate~\cite{keller2016geometrical,steinberg2024free}. We employ NVIDIA Sionna RT~\cite{sionna}, which models reflection, refraction and diffraction. Though far from a full-wave solver, we find it quite effective. Our Sionna wrapper probes the scene at frequencies relevant to the selected radar, e.g., at a discrete sampling interval across 77--81\,GHz for a 77\,GHz waveform with 4\,GHz bandwidth. We also deselect the direct Tx-to-Rx line-of-sight path, which radars are designed to minimize. Sionna models amplitude and phase, but the measured phasors, aka Channel Frequency Response (CFR), are not yet what the radar sees.

To bridge this gap, we perform sensor simulation with a GNU-radio-style breakdown of radar sub-components, one per radar hardware in our database, accounting for these prominent parameters: \texttt{antenna patterns}, \texttt{waveform type}, \texttt{waveform parameters}, \texttt{waveform repetition structure}, \texttt{ADC rate}, and \texttt{effective noise floor}. Spatial features like \texttt{antenna array structure} are modeled in the world simulator. The Sionna wrapper followed by the GNU-radio-like signal chain forms our simulator, generating digital radar signals.

\begin{figure}[!htb]
\centering
\vspace*{-0.3cm}
\includegraphics[width=0.8\columnwidth]{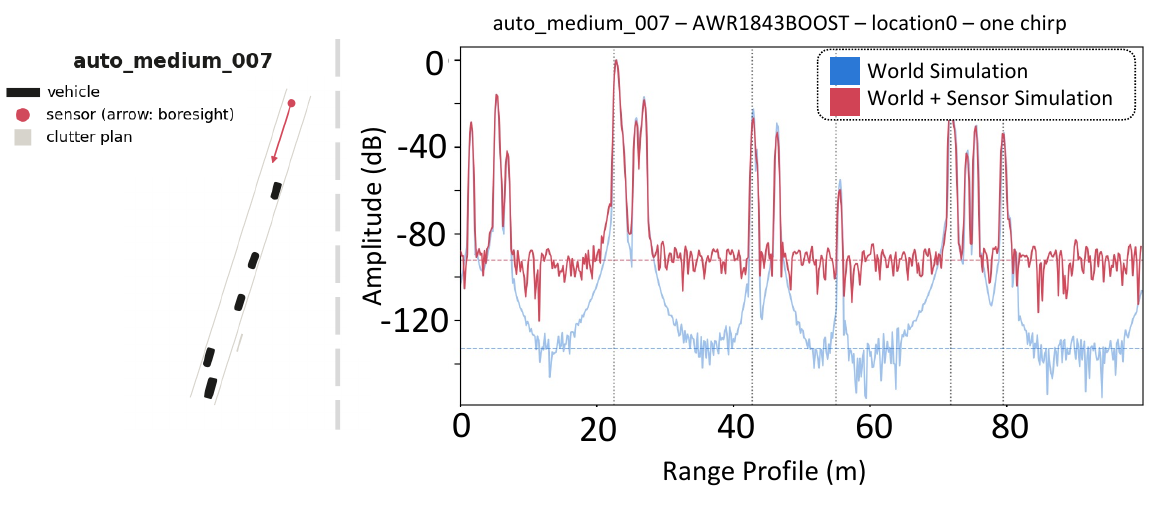}
\vspace*{-0.2in}
\caption{\name\ performs RF and sensor simulation to generate realistic digital signals. This figure shows the increased noise floor in range profile after sensor simulation.}
\vspace*{-0.15in}
\label{fig:sensor_sim}
\end{figure}

\sssec{On-Sensor Compute Emulation.} Accurate on-sensor emulation matters because user constraints are often compute-related: time to task completion, power budget, and memory budget. Radar modules in our database are user-programmable, but narrowly, as on-sensor compute is limited. For example, the popular AWR1843 radar has 512\,kB program RAM and 128\,kB data RAM on chip, insufficient for long-horizon frame tracking, frame aggregation and iterative algorithms needing more memory. Sensor modules are therefore almost always paired with off-sensor compute for advanced processing. On-sensor coding blocks reduce to multi-stream demultiplexing, range-angle-Doppler processing, thresholding, raw data streaming, or a combination, and vendors benchmark the enabling routines (\texttt{fft128, fft256,} etc.) in datasheets. Hence, we emulate on-sensor performance rather than execute on a real sensor.

Our emulator is a three-stage, trace-driven pipeline: a twin of the radar frontend, on-device processing, and output transfer. Stages produce or consume timed events, which a discrete-event simulator (DES) composes to capture steady-state behavior and transients.

\textbullet\hspace*{0.1cm} The frontend checks the radar configuration against device capabilities, then synthesizes frame-level ADC samples of requested dimensions and timing.

\textbullet\hspace*{0.1cm} The processing stage runs On-Sensor SA generated C code through a per-frame callback. The interface supports batching and temporal processing while exposing timing, memory use, and output bursts. We compile C to the target ISA and model each basic block with \texttt{llvm-mca}, weighted by measured execution counts. For operations invisible to \texttt{llvm-mca}, instrumented wrappers intercept library calls (e.g., \texttt{malloc}, \texttt{memcpy}, \texttt{sqrt}) and stubs record device-specific calls (e.g., DSP FFTs, hardware-accelerator operations), with costs from vendor documentation or benchmarks. For memory, \texttt{llvm-size} measures static RAM and program size, Clang's \texttt{-fstack-usage} estimates peak stack, and intercepted \texttt{malloc}/\texttt{free} calls track peak heap. We check static data, stack, and heap against RAM capacity, and program size against flash capacity.

\textbullet\hspace*{0.1cm}The output stage sends processed results, or raw samples if processing is bypassed, over the configured link (e.g., Ethernet, SPI, UART), tracing transfer rate, buffering, and burst timing.

From all three traces, the emulator checks configuration legality, frontend-to-processor buffer capacity, real-time compute feasibility, RAM/flash capacity, and link throughput/latency, and gives a confidence score.

\sssec{Off-Sensor Execution.} To allow a broad scope of algorithms (from real-time multi-frame tracking to iterative imaging workloads) on edge devices like a Raspberry Pi, NVIDIA Jetson or high-end NVIDIA GPU, we simply build a rig of these devices and execute code for real, verifying that it runs and estimating its time-to-completion and memory. The Off-Sensor Software SA's last step is rule-based selection of edge-device power-mode settings from user requests. A dispatcher maintains a virtual queue. Once the appropriate edge device is available, it dispatches the first queued job, clears the background and sets the power mode (\texttt{scaling-governor} for RPi, \texttt{nvpmodel} for NVIDIA GPUs). We measure execution time with the target device's timer, excluding startup, data I/O, and warm-up, sample power through INA3221 on the Jetson and \texttt{vcgencmd} on the Raspberry Pi, and enforce memory limits through \texttt{systemd} groups.

\subsection{Efficient Agent Interaction}
\label{ssec:fast}

Simulation and execution take time (Fig~\ref{fig:time}, single-prompt output), so we embrace all opportunities to enhance efficiency.

\subsubsection{\underline{Efficiency in evaluation}}

\begin{figure}
\centering
\includegraphics[width=\columnwidth]{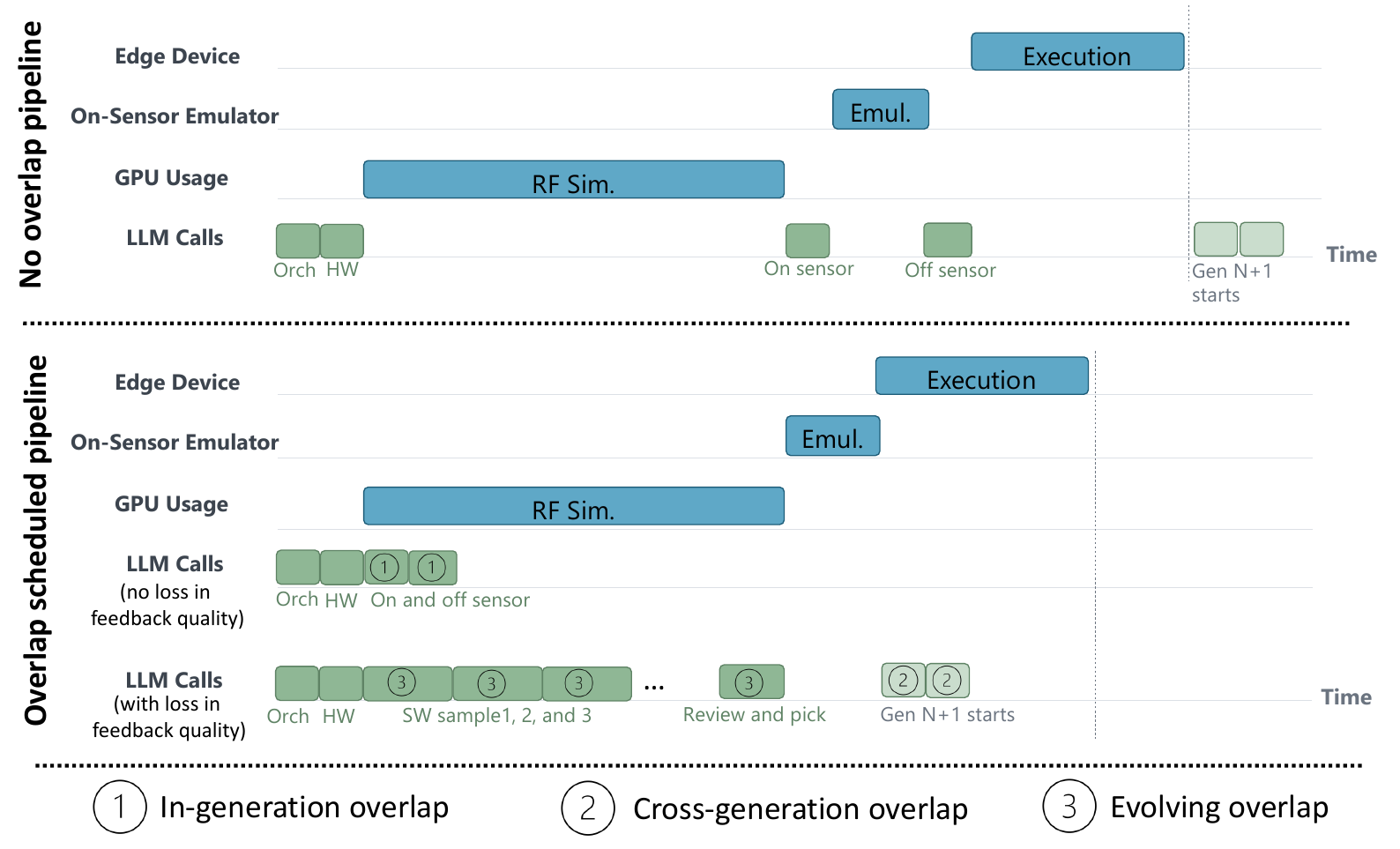}
\vspace*{-0.3in}
\caption{An illustration of overlap scheduling of GPU-intensive tasks with LLM queries. 
This assumes no failures from local verifiers in each SA, no \emph{Optimize} LLM, and no \emph{Param sweep} LLM for simplicity.}
\vspace*{-0.2in}
\label{fig:efficiency}
\end{figure}

\sssec{Overlap scheduling.} A naive implementation is sequential (Fig~\ref{fig:efficiency}), and we provide three modes to improve its latency. First, \emph{in-generation overlap} runs software SAs' cloud LLM queries concurrently with the current generation's RF simulation, the agent's GPU-intensive step. The next two modes further improve latency at the cost of accurate feedback. \emph{Cross-generation overlap} starts the next generation while the current one is still under evaluation, overlapping generation $N\text{-}1$'s compute stages with $N$'s \emph{Orchestrator} or hardware SA calls, and $N\text{-}1$ is unscored in $N$'s LLM calls. \emph{Evolving overlap} uses the long RF simulation time to generate multiple candidates, from which an LLM reviewer selects the most promising one, using idle LLM capacity to probe a larger search space and improve the solution submitted for evaluation. Users choose the mode upfront, preferring accurate feedback or a larger search space. Sec~\ref{ssec:eval_4} evaluates each mode's effect.

\sssec{Memory-efficient tiled CFR synthesis.} Our main CFR optimization avoids the large broadcast intermediates in Sionna's path-to-frequency conversion. For antenna pair $i$, chirp $n$, and frequency offset $\Delta f_k$, the CFR is
\begin{equation}
H_{i,n,k}=\sum_p a_{i,p}
 e^{-j2\pi f_0\tau_{i,p}}
 e^{j2\pi\nu_{i,p}t_n}
 e^{-j2\pi\Delta f_k\tau_{i,p}},
\label{eq:cfr-path-sum}
\end{equation}
where $a_{i,p}$, $\tau_{i,p}$, and $\nu_{i,p}$ are the gain, delay, and Doppler shift of path $p$, and $f_0$ is the center frequency. The broadcast formulation materializes every path's contribution at every antenna pair, chirp, and frequency, an $I\times P\times T\times F$ intermediate, only to sum over $P$ into an $I\times T\times F$ CFR. We instead evaluate Equation~\ref{eq:cfr-path-sum} in tiles with fused Triton GPU kernels. For antenna pair $i$, let $\mathcal N$, $\mathcal P$, and $\mathcal K$ select blocks of chirps, paths, and frequencies, accumulating into a zero-initialized $H$:
\begin{equation}
\begin{aligned}
A_{i,n,p}&=a_{i,p}e^{-j2\pi f_0\tau_{i,p}}
 e^{j2\pi\nu_{i,p}t_n},\\
B_{i,p,k}&=e^{-j2\pi\Delta f_k\tau_{i,p}},\\
H_{i,\mathcal N,\mathcal K}&\leftarrow H_{i,\mathcal N,\mathcal K}
 +A_{i,\mathcal N,\mathcal P}B_{i,\mathcal P,\mathcal K}.
\end{aligned}
\label{eq:cfr-tiles}
\end{equation}
Each GPU program generates $A$ and $B$ tiles on chip and immediately accumulates their product over the current path block, so only partial CFR sums reach global memory and the path-expanded tensor is never stored. The frequency-phase tile $B$ is reused across all chirps, and any chirp count, down to one, is handled. Tile sizes and frequency batches adapt to available GPU resources without changing results beyond floating-point rounding. We also reuse compiled kernels across sensor placements, chirp configurations, and scene frames to avoid recompilation. On an NVIDIA RTX 5090, these optimizations speed up simulation by $13.9\times$ for multiple sensor placements, $128.4\times$ for multi-chirp acquisitions across placements, and $164.2\times$ for sequential frames in time-varying scenes.

\sssec{Progressively-scaled scene difficulty.} With a ray-tracer, scene complexity directly dictates RF simulation time. For example, large diffuse surfaces like a floor or substrate contribute at every radar position, and more received rays mean longer simulation. Object count, sizes, and scattering properties all define complexity. Test engineering teams avoid unnecessary tests with good testing principles. For example, an automotive radar is rigorously field-tested first with a pair of corner reflectors, then on a motor track with controlled obstacles, and eventually on unconstrained streets. We observe that stack layers generated in initial agent iterations perform fairly weakly and need not be tried on complicated scenes, so we scale scene complexity progressively, based only on successful past performance on simpler scenes. We label scenes easy, medium or hard manually, or by rules for automatically generated ones. Central SA raises difficulty by rule: a new tier unlocks when any archive entry achieves a median score, across current scenes, of 90\% of the requirement's target score, at most one per generation.

\subsubsection{\underline{Efficiency due to agent structure}}

\sssec{Parameter sweeper.} Radar algorithms need carefully chosen parameters, the best known being CFAR (Constant False Alarm Rate), a threshold that converts heatmaps into point clouds near the end of processing chains. Other thresholds lie at intermediate steps. For example, mmNorm performs Raw Data $\rightarrow$ Backprojection $\rightarrow$ Normal Estimation $\rightarrow$ Segmentation $\rightarrow$ Isosurface optimization $\rightarrow$ Output point cloud, with thresholds chosen at each step. Thus, if the Off-Sensor Software LLM misses the perfect parameter set at first, it repeats all steps. Empirically, LLMs mostly know the general steps but struggle to find the right parameters, and new agent iterations are too slow for parameter selection. We instead use (1) a cache storing unthresholded outputs at each step of the end-to-end evaluator, and (2) a \emph{Parameter Sweep} LLM that proposes new parameters from cached results and evaluator feedback, within one agent iteration, i.e., a local rather than cross-iteration search.

\sssec{Edge-aware optimization.} SA-generated code may be functionally correct but computationally inefficient, particularly in the Off-Sensor SA, which can propose complicated, time-consuming algorithms. We include an Optimizer SA call before real execution on the edge device. Under the user-requested compute constraints, it produces hardware-aware implementations with frameworks such as Triton, JAX, or \texttt{torch.compile}, preserving the original algorithm. This saves time otherwise wasted on executions that eventually fail the user's edge-compute requests.

\sssec{Cross-generation caching of intermediate outputs.} As evaluation is expensive, we cache at two levels. The \textit{first} reuses outputs of unchanged stack layers across generations. E.g, when an agent modifies only the off-sensor code, configuration, RF simulation, and on-sensor computation are reused. When the full simulation cannot be reused, the \textit{second} operates per sensor placement and configuration, so only new or modified ones are simulated, maximizing reuse of computed data and avoiding wasted time.

%% file: evaluation.tex
\section{Implementation}
\label{sec:impl}

\name\ is implemented in Python and orchestrates LLM APIs served on Google Enterprise Agent Platform (Gemini and third-party models such as Claude). By default, the primary and lightweight LLMs are Gemini 4 Argon and Gemini 3.8 Flash with ``high'' thinking (Fig~\ref{fig:overview}). Simulation and evaluation run on an AMD Ryzen Threadripper PRO CPU with an NVIDIA RTX A5000 GPU, and off-sensor (L5) code runs on the user's target platform: an RTX A5000, Jetson Orin, or Raspberry Pi 5. For radar selection and emulation, we support 10 radars/radios and companion boards across sub-10, 24, 60, and 77\,GHz bands: AWR1843\allowbreak BOOST, AWR2243-\allowbreak 2X-\allowbreak CAS-\allowbreak EVM, AWR6843\allowbreak AOPEVM, MMWCAS-\allowbreak RF-\allowbreak EVM, DEMO BGT60\allowbreak TR13C, DEMO BGT24\allowbreak ATR22 YPA, OPS241-\allowbreak B-\allowbreak FM-\allowbreak RP, DWM3001\allowbreak CDK, XeThru X4, and LimeSDR Mini 2.0.

We compare against one-shot prompting, OpenEvolve~\cite{openevolve}, and Claude Code, whose general-purpose frameworks best suit us. Recent agent papers often omit Claude Code, but its recent progress makes it an essential baseline. Each baseline gets the same user requirements, task documentation, and evaluation protocol as \name. One-shot prompting and OpenEvolve use Gemini 4 Argon with high effort, and we strive for a functional single-prompt solution. For OpenEvolve, a meta-prompt asks it to write a deterministic, task-specific evaluator scoring in $[0,1]$. OpenEvolve then refines and selects solutions by this score (70\%) and its native LLM-as-a-judge feedback (30\%). Claude Code (\texttt{claude -p}, Opus 5.5, high effort, 24-hour limit) is informed of our coding abstractions and asked to design all five layers in one session, sandboxed to the task materials and designated working directories. Its built-in \texttt{Bash}, \texttt{Read}, \texttt{Edit}, \texttt{Write}, and \texttt{Agent} tools let it profile the machine, write and execute code, and delegate to subagents.

We also evaluate two hybrids. (1) \emph{Claude Code with \name\ tools} additionally receives six \name\ tools: \texttt{simulate} generates radar measurements, \texttt{run\_kernel} compiles and tests sensor kernels, \texttt{emulate} checks hardware feasibility, \texttt{run\_analysis} executes the off-sensor pipeline, \texttt{run\_on\_device} runs it on the target edge device, and \texttt{score} measures task performance. Six skills, derived from \name's agent prompts and task package, distill its design principles: \emph{five-layer-stack} and \emph{shape-as-code} (layered design and representations), \emph{review-and-pictures} and \emph{search-memory} (visual review and design-archive reuse), \emph{sweep-and-optimize} (parameter tuning and runtime optimization), and \emph{subagents} (delegation). Claude Code decides when and how to use them and may delegate to up to four subagents, a decomposition comparable to \name's. This separates the benefit of \name's tools and insights from that of its orchestration harness. (2) \emph{\name\ with coding agents} keeps \name's orchestration and layered structure but replaces each layer-specific LLM call with a Claude Code session that can edit code, write scripts, and run analysis on the execution backends. To approximate a single primary-LLM call, each session uses medium effort, a 45-minute timeout, and no \texttt{Agent} tool. This tests whether general-purpose coding complements structured orchestration.

Average costs per design request (input/output tokens) are \$0.92 for one-shot prompting (0.21M/0.06M), \$10.58 for OpenEvolve (3.33M / 0.60M), \$11.08 for Claude Code (22.70M/0.18M), \$16.37 for Claude Code with tools (48.13M/0.17M), \$77.16 for \name\ for 20 iterations (25.49M/5.24M), and \$98.00 for \name\ with coding agents (100.47M/1.44M). Iterative design drives both \name\ variants' higher cost, still orders of magnitude below what organizations spend developing a radar solution. \name's median running time is 17.9 hours. API calls cost $\approx\$10{,}000$ in total, including early experiments and failures. We tested \name, its ablations, and baselines on 362 prompts.

\begin{figure*}[t]
\centering
\includegraphics[width=\textwidth]{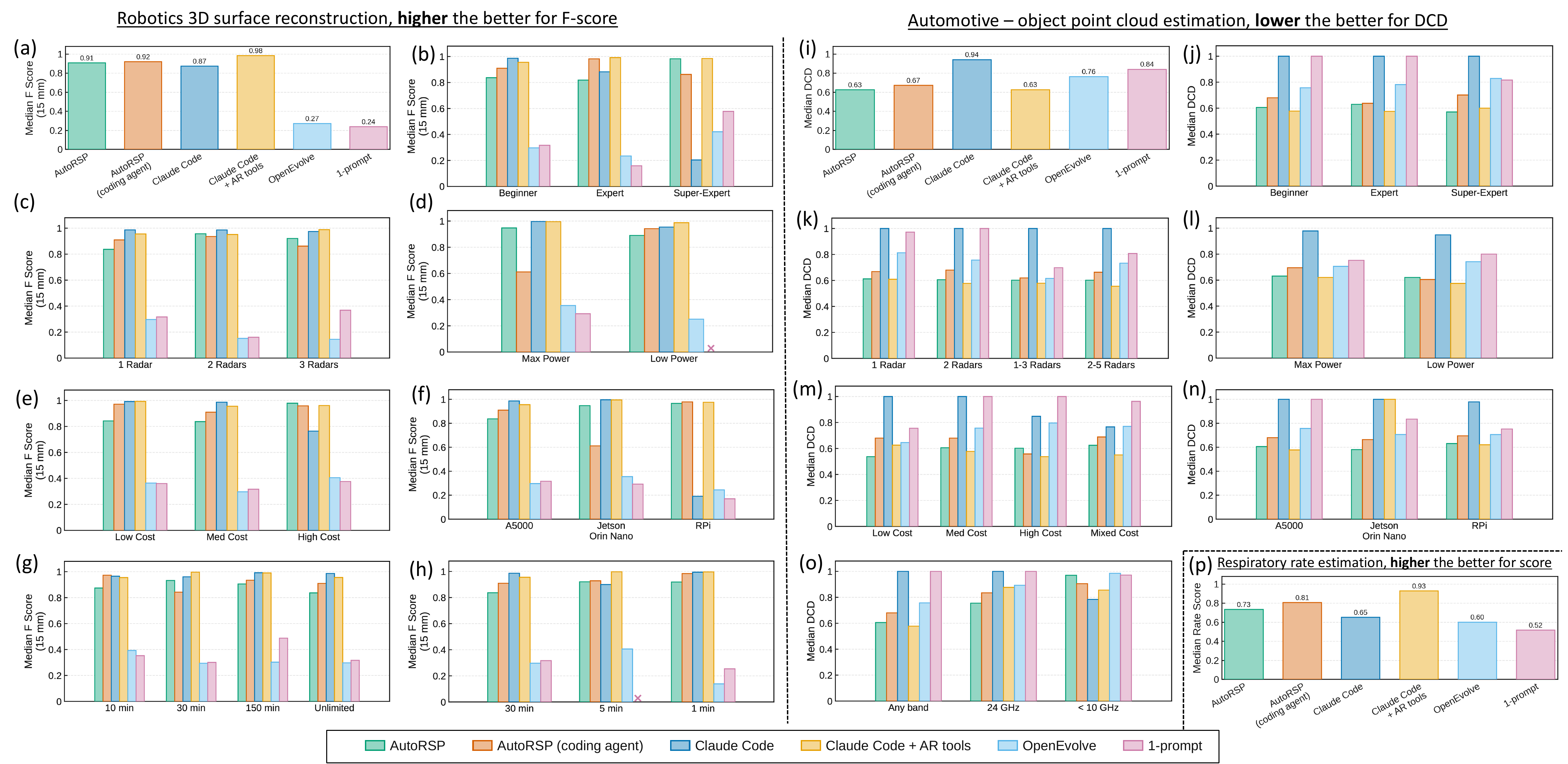}
\caption{Across robotic 3D surface reconstruction (a--h), automotive point-cloud estimation (i--o), and respiratory-rate estimation (p), we compare \name\ (green), \name\ with coding agents (orange), Claude Code (blue), Claude Code with \name\ tools (yellow; ``Claude Code + AR tools'' in the legend), OpenEvolve (light blue), and one-shot prompting (pink). Panels (a,i,p) summarize task performance. We vary user expertise (b,j), radar count and placement-area constraints (c,k), edge-device power budgets (d,l), radar hardware cost (e,m), target edge device (f,n), acquisition time (g), edge execution-time budget (h), and radar frequency band (o). Higher F-score and respiratory-rate score are better; lower Density-aware Chamfer Distance (DCD) is better.}
\vspace*{-0.3cm}
\label{fig:main_results}
\end{figure*}

\section{Evaluation and Results}
\label{sec:eval}

\subsection{Across User Requests}
\label{ssec:eval_1}

\sssec{Methodology.} We benchmark LLMs and agentic systems on three RF sensing tasks: robotic 3D surface reconstruction~\cite{NLOS}, automotive object point-cloud estimation~\cite{bansal2020pointillism}, and respiratory-rate estimation during exercise~\cite{msense}. We use 50 prompts (20, 25, and 5, respectively) that vary user expertise, SWaP-C constraints, deployment platforms, frequency bands, and acquisition and execution time budgets. We run \name\ for up to 20 agent iterations / generations per request. Robotic imaging is scored by F-score at a 15\,mm threshold~\cite{NLOS} and automotive by Density-aware Chamfer Distance~\cite{dcd}. Vital sensing is scored by $s_{\mathrm{resp}} = 1 - \min(\mathrm{MAE}, 10)/10$, where MAE is the mean absolute error in breaths/min between estimated and ground-truth respiratory rates at matching timestamps. Every solution proposed is checked against the specified SWaP-C constraints, assessed for feasibility by our on-sensor-emulator, and tested for execution on the target edge device within the requested budget. Solutions that fail any of these checks receive the worst possible score.

\begin{figure*}[t]
\centering
\includegraphics[width=\textwidth]{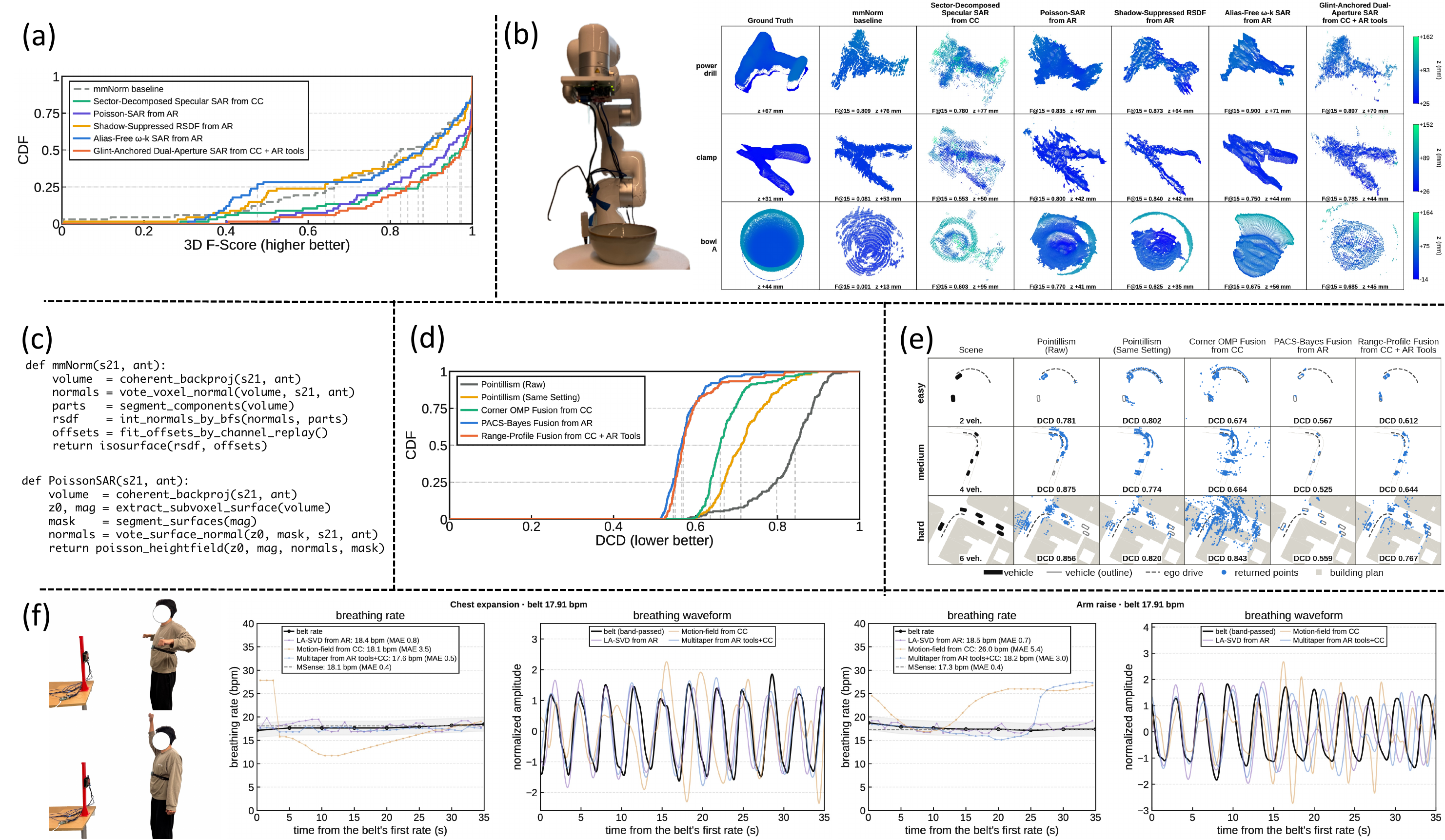}
\caption{Deep dives (AR = \name, CC = Claude Code): (a-c) show \name\ against mmNorm for the robotic task. (a) shows Poisson-SAR named and proposed by \name\ to outperform mmNorm. (b) shows qualitative results on real objects. Note that estimating the correct median Z is as important as visually appearing like the ground truth. mmNorm wrongly estimates in several cases and with a 15 mm threshold for F-score gets penalized. (c) shows a preview of L5 approach of Poisson-SAR. (d-e) show \name's superior performance against Pointillism. Qualitatively, we are showing a point cloud aggregation over a short drive for intuitive comparison. Observe the superior point placement on true car outline in \name\ vs Pointillism. (f) shows real world testing on respiration estimation where \name\ achieves comparable performance to a leading human-expert-baseline, MSense.}
\vspace*{-0.3cm}
\label{fig:deep_dive}
\end{figure*}

Scenes are labeled easy, medium, or hard, either manually or by predefined rules for generated scenes. The robotic imaging task uses YCB dataset, ranging from single flat-surfaced objects to curved or complex objects to very small or multi-surface objects. For automotive, we create 150 scenes of 25 frames each, ranging from vehicles on flat roads to roads with curbs and additional objects to Sionna urban environments modeled on Munich, Paris, and Florence. For vital sensing, a coding agent builds and refines 150 scenes of MSense-like exercises~\cite{msense} from body assets and motion templates, with physiologically plausible breathing displacements and rates~\cite{vital_sign_displacement,vital_sign_bpm}. Easy scenes have stationary people breathing deeply at a fixed rate. Medium scenes add exercises, such as chest expansion and side bends, that keep the chest roughly within one range--angle cell. Hard scenes combine shallower breathing with chest motion across cells or arm occlusion (lunges, squats, front arm raises), plus walking bystanders and possibly changing respiratory rates. From these, 8 to 10 randomly chosen scenes guide \name's search, innovation and evaluation, and 20 randomly selected held-out-test scenes assess how well the methods proposed by \name\ generalize to unseen scenes.

\sssec{Across tasks.}
\name\ outperforms one-shot prompting, Open\allowbreak{}Evolve, and Claude Code, none of which use our tools or harness (aggregate results in Fig~\ref{fig:main_results}a, i, p). Claude Code builds and iterates against its own simplified simulator, which keeps its code executable. It still trails \name\ and Claude Code with tools, which get grounded, hardware-aware feedback from our tools. With \name's six tools and six skills, Claude Code with tools matches or exceeds \name\ on all three tasks, so our tools and insights carry over to other agent harnesses.

In the robotic imaging task (Fig~\ref{fig:main_results}a), the mean of the per-scene median F-scores are 0.98 for Claude Code with tools, 0.92 for \name\ with coding agents, 0.91 for \name, and 0.87 for Claude Code, OpenEvolve (0.27) and one-shot prompting (0.24). Raw data can usually be streamed off-sensor within hardware and time constraints, so emulation failures are rare and Claude Code loses less ground here than automotive or vital signs estimation tasks (also seen in Fig~\ref{fig:main_results}b-h).

In the automotive task (Fig~\ref{fig:main_results}i), \name\ reaches an average of median DCD of 0.63 (lower is better), versus 0.76 for OpenEvolve and 0.84 for one-shot prompting. The baselines rely on established methods from the LLM's prior knowledge, producing functional but poorly optimized solutions. Real-time requirements force most processing on-sensor, and without an emulator Claude Code judges feasibility by reasoning alone. It treats each layer as a hard coding problem, pushes increasingly complex code onto the sensor, and often fails emulation. Even with unlimited on-sensor compute, it still trails \name, and Claude Code with tools.

In respiratory-rate estimation (Fig~\ref{fig:main_results}p), median scores averaged across prompts are 0.93 for Claude Code with tools, 0.81 for \name\ with coding agents, 0.73 for \name, 0.65 for Claude Code, 0.60 for OpenEvolve, and 0.52 for one-shot prompting. 

We attribute Claude Code's weaker performance on automotive and vital sensing tasks to lack of grounded evaluation. Simulating scenes for robotic imaging is simpler (i.e; Claude Code could just fetch primitives from trimesh) than complex human exercise motion or car environments. Across all 50 Claude Code designs, the self-written simulators follow an easy approach, modeling reflections as coherent sums of single-bounce signals directly from points sampled on simplified geometries. They rely on manually specified reflectivity instead of material-dependent properties and generally neglect diffraction and multi-bounce reflections. On-sensor and off-sensor constraints further make it harder to judge. Next, we look at the impact of one-variable change in user request.

\sssec{Across user expertise.}
Novice prompts give only the requirements imposed by Spec-SA, expert prompts add a pseudo-algorithm, and super-expert prompts give an exact implementation. \name\ handles all three (Figs~\ref{fig:main_results}b, j) and improves with more guidance: from novice to super-expert, robotics F-score rises from 0.84 to 0.98 and automotive DCD improves from 0.61 to 0.57.

\sssec{Across size, cost, and power.} Radar-cost budgets are \$200--\$500, \$1,200, and \$5,000, and size, weight, and placement area follow platform constraints. Fig~\ref{fig:main_results}c--e and k--m show \name's performance across radar count, hardware price, and edge-device power budget. High-end systems stream raw measurements through a dedicated capture board (e.g., an AWR1843 with a DCA1000), which the low-cost budget cannot afford. \name\ instead generates DSP-targeted C kernels that keep only the top-$K$ CFAR range detections, refined to sub-bin precision, along with their 12-channel complex IQ samples. It then builds downstream software around this compact output and still achieves reasonable task performance. The coding-agent variant's dip occurs due to a one-off erroneous interpretation of digital radar signals that somehow still works on flat objects raising the \textit{Reviewer's} optimism falsely.

\sssec{Across edge devices.}
We evaluate on widely available edge devices with and without GPUs: a Jetson Orin Nano in 15\,W and \texttt{MAXN\_SUPER} modes, and a Raspberry Pi at 1.5 and 2.4\,GHz. \name\ adapts its code to each target (Fig~\ref{fig:main_results}f, n), using OpenMP parallelization on the RPi and Triton GPU acceleration on Jetson.

\sssec{Across collection time and execution time.}
Fig~\ref{fig:main_results}g, h show that \name supports different acquisition and execution time budgets. Using robotic-arm speed parameters based on a real arm, scans under 30 minutes collect fewer than 2,000 frames, below the Nyquist sampling requirement. \name\ with coding agents performs best under the 10-minute scan budget by combining dense overhead sampling with outer viewpoints and varied board orientations to capture tilted, specular facets. Its reconstruction also accounts for this.

\sssec{Across frequency bands.}
Restricting radar selection to 24\,GHz or sub-10\,GHz reduces automotive sensing performance (Fig~\ref{fig:main_results}o). Our supported sub-10\,GHz devices lack the sensing range for this task, and the 24\,GHz radars' lower bandwidth limits range resolution. Without restrictions, all methods converge to mmWave radars.

\subsection{Deep dive to contrast against human-experts}
\label{ssec:eval_3}

\sssec{Methodology.} We next present results from an expert prompt designed to push \name\ beyond human-engineered solutions. We compare the generated radar stacks against mmNorm~\cite{NLOS} and Pointillism~\cite{bansal2020pointillism}, which handle specular targets through algorithmic processing and additional hardware, respectively. For respiratory-rate estimation, we compare against MSense~\cite{msense}. mmNorm runs its original pipeline and codebase on our simulated measurements, with its backprojection kernel reimplemented in Triton for speed without changing its output. mmNorm and \name's solution receive the same measurements and backprojection volumes. For Pointillism, we reimplement the TI mmWave processing pipeline in Python from the TI Radar Toolbox~\cite{ti_radar_toolbox_scene} and refine its point cloud with a cross-potential filter. MSense is not publicly available, so we reimplement it in Python from the paper and manually tune its thresholds on real-world data, collected under IRB approval.

\begin{figure*}[t]
\centering
\vspace*{-0.3cm}
\includegraphics[width=\textwidth]{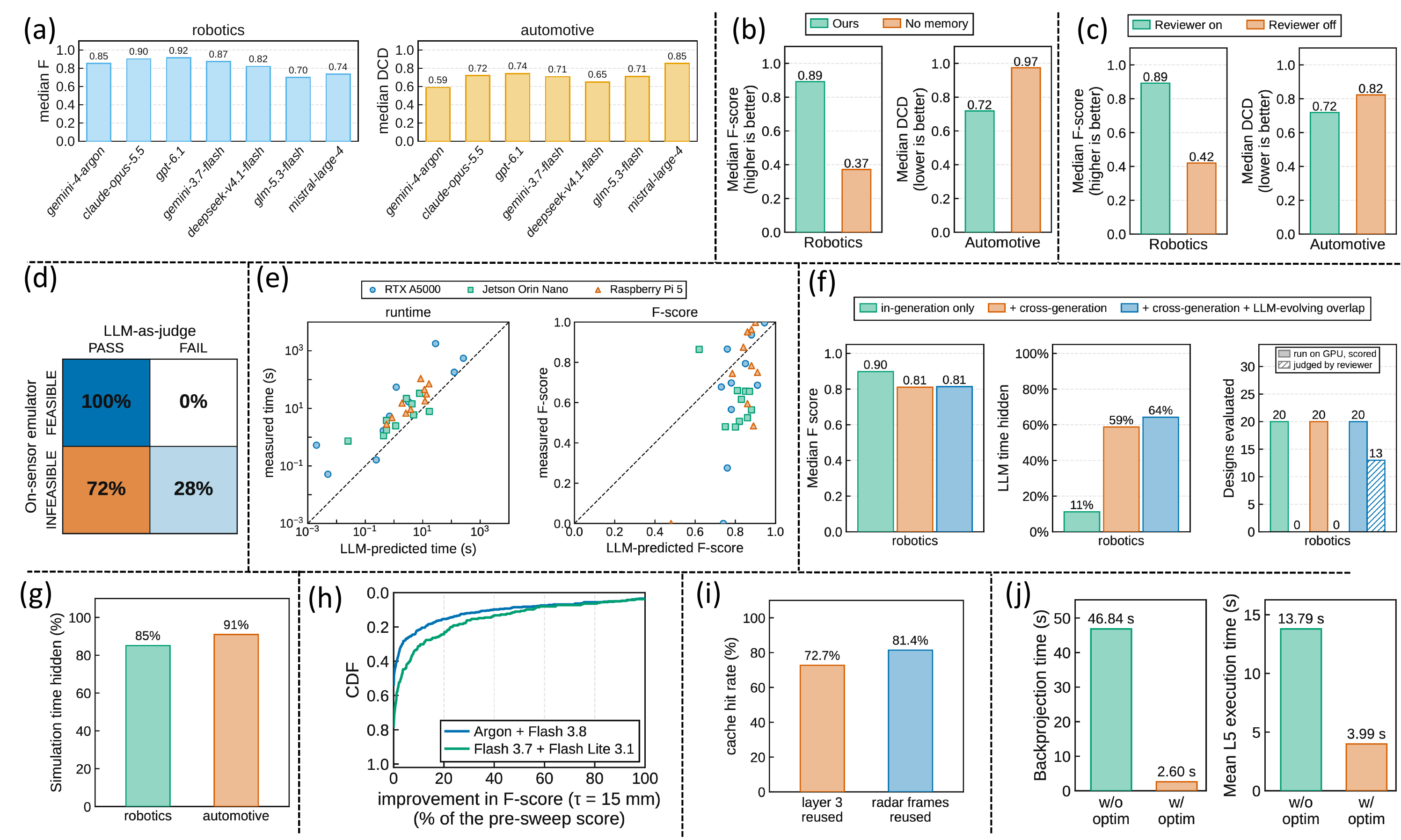}
\vspace*{-0.7cm}
\caption{Ablation evaluations of \name. (a) Model comparison, reporting the mean across three prompts of each run's median held-out-test scene score. (b) Effect of memory structure. (c) Effect of the reviewer and image-based feedback. (d) Confusion matrix comparing LLM-judge feasibility decisions with \name's sensor emulator. (e) LLM-predicted versus measured execution time and F-score; dashed lines indicate perfect agreement. (f) Effect of scheduling strategies on sensing performance, hidden LLM time, and the number of evaluated designs. (g) Simulation time hidden by in-generation overlap for Argon. (h) F-score improvements from a lightweight parameter sweeper. (i) Layer-level and radar-frame cache reuse. (j) Execution time before and after optimization for a representative backprojection kernel (left) and averaged across 20 selected generations (right). Higher F-score and lower DCD indicate better performance.}
\vspace*{-0.3cm}
\label{fig:ablations}
\end{figure*}

\sssec{Robotic imaging.}
Across all 67 YCB scenes, several of \name's solutions match or exceed mmNorm's F-score (Fig~\ref{fig:deep_dive}a) in simulation. Claude Code also generates competitive approaches, but \name\ and Claude Code with \name's tools have better tail performance. The compared designs explore a few distinct ideas. \name's Poisson-SAR (Fig~\ref{fig:deep_dive}c) is simple but effective: it smooths peak heights while anchoring them to strong returns. Its Shadow-Suppressed RSDF builds on normal-based reconstruction, adding primary-lobe isolation and shadow suppression to reject spurious returns. Claude Code's Sector-Decomposed Specular SAR reconstructs each viewing direction separately before merging them, which captures more facets but also admits false returns. This shows how \name\ and its tools can explore early-stage research ideas and expose their strengths and limitations.

For real-world validation, we mount an AWR1843BOOST radar on a UFACTORY xArm 6 robotic arm, use a RealSense D435 depth camera for ground truth, and image a clamp, a power drill, and a bowl. On flat objects, \name's improvements are marginal (Fig~\ref{fig:deep_dive}b). mmNorm generally reconstructs flat surfaces well, but its sampling of 50 receive channels and its isosurface optimization can select different surfaces across runs, causing score variability. It is less reliable on hollow objects, where returns may come from multiple surfaces. On the bowl, mmNorm collapses the geometry into a single, roughly flat surface, whereas \name's solution recovers a smoothed bowl profile, including parts of its bottom and curved sides. Worth noting is, unlike the catastrophic one-layer result from Fig~\ref{fig:motivation_v2}, merely asking Claude Code to use our abstractions of 5 layers has already resulted in better performance.

\sssec{Automotive.} Median DCD over 150 scenes falls from 0.84 with Pointillism's published setup to 0.71 with Pointillism processing on \name's measurements, and to 0.57 with \name's full stack (Fig~\ref{fig:deep_dive}d), so \name\ improves both configuration and processing. Qualitative results are shown in Fig~\ref{fig:deep_dive}e. It picks two 62\,GHz AWR6843AOP radars over two 77\,GHz AWR1843BOOSTs for wider elevation coverage, and extends the unambiguous range from 20\,m (set by Pointillism's sampling parameters, not the hardware) to 107\,m. On-sensor, separate stationary and moving integrations with multi-scale CFAR replace a single host-side range--Doppler CFAR. Off-sensor, \name\ refines angles with all 12 virtual channels and downweights radar-fixed clutter, such as under-bumper road returns, unless the other radar corroborates it. These changes keep weak vehicle returns while suppressing spurious points. Claude Code with \name's tools also outperforms Pointillism. Standalone Claude Code detects cars in simple scenes but struggles in cluttered environments with buildings and other objects.

\sssec{Vital-signs sensing.}
We collect real measurements with a TI AWR1843 BOOST radar at a 1\,ms frame interval, following MSense's setup~\cite{msense}, and record ground-truth respiration with a Go Direct Respiration Belt~\cite{vernier_respiration_belt} (Fig~\ref{fig:deep_dive}f). On the chest-expansion and arm-raise recordings, our MSense implementation achieves MAEs of 0.4 breaths/min on both, and \name's best method, LA-SVD, approaches this at 0.8 and 0.7, while standalone Claude Code trails at 3.5 and 5.4. These results suggest \name\ can help produce open alternatives to published methods with comparable performance.

\subsection{Ablation Study}
\label{ssec:eval_4}

\sssec{Methodology.} We ablate (1) performance across base LLMs, (2) performance without the stack-awareness and evaluation feedback (Sec~\ref{ssec:physics-eval}), and (3) effectiveness of efficiency methods in Sec~\ref{ssec:fast}.

\subsubsection{Model Choice}
We evaluate seven model configurations. Four pair a primary reasoning model with a lightweight one for auxiliary tasks to cut cost: Gemini 4 Argon/3.8 Flash, Claude Opus/Sonnet 5.5, GPT-6.1 Sol/5.6 Terra, and Gemini 3.7 Flash/3.1 Flash Lite. Three use a single open-source model for all agents via OpenCode~\cite{opencode}: DeepSeek V4.1 Flash, GLM 5.3 Flash, and Mistral Large 4. Each configuration runs three prompts per task for 20 generations on \name's design-time scenes, and we report the mean, over prompts, of each run's median held-out-test scene score. Closed-source frontier models score highest, with Gemini 4 Argon slightly ahead in automotive sensing (Fig~\ref{fig:ablations}a). Open-source models also run end to end, so \name's LLM backbone is interchangeable, though sensing performance tracks model capability.

\subsubsection{Intermediate Structure}
\sssec{Memory structure.}
Without memory, \name\ cannot reliably solve complex radar-design tasks (Fig~\ref{fig:ablations}b). Layer decomposition eases each decision, but the agent rarely gets every layer right within a few attempts. The memory (\textit{Archive}) retains past designs and outcomes, so agents learn within layers and across layer dependencies, progressively improving the full sensing stack.

\sssec{Reviewer.}
Removing the \textit{Reviewer} tests the value of explicit cross-layer feedback. The orchestrator must then inspect the stack, infer each subagent's past contribution, and give layer-specific feedback itself, and it performs consistently worse on robotic imaging and automotive tasks (Fig~\ref{fig:ablations}c). In robotic imaging, its score drops to 0.42, and it fails to reconstruct shapes even in the $(x,y)$ plane, because missing cross-layer feedback limits hardware-software co-design.

\subsubsection{Accurate Feedback}
\sssec{Emulator vs LLM as feedback.} 
We first benchmark our emulator against true time and memory measurements on programs flashed on AWR1843BOOST. We find that the mean absolute error is 1.3\% of the frame time and 64 bytes for memory. Next, we select 60 unique radar-board C code snippets from \name-generated solutions: 10 from robotic imaging tasks, which mostly stream raw data with limited on-sensor processing, and 50 from automotive tasks, which stress on-sensor computation. As an LLM judge, Gemini 4 Argon (high effort) disagrees with the emulator (Fig~\ref{fig:ablations}d), tending to accept code it explicitly flags as infeasible under the board's constraints. This shows the limits of LLM-only assessment for physical systems.

\sssec{Real edge device vs LLM as judge.}
Fig~\ref{fig:ablations}e compares LLM predictions for 30 off-sensor programs spanning a range of runtimes and task scores. Using the emulator judge's model and reasoning settings with separate prompts, the LLM estimates each program's runtime and task performance, which we compare against measured runtime on the target device and the evaluation pipeline's score. The LLM captures the relative runtime trend but underestimates runtime by $4.1\times$ at the median. For e.g, it predicts 28.0\,s for the longest-running program, which takes 1778.5\,s on the RTX A5000, so programs judged within the user's time budget can violate it in deployment. For task performance, it gives nearly constant, overestimated scores, often anchored to the prompt's target value rather than the program's actual output. LLM-only judging thus captures neither hardware-dependent runtime nor actual task performance.

\subsubsection{Efficiency Feedback}
\sssec{Efficiency Modes}
\label{ssec:overlap-efficiency}
Fig~\ref{fig:ablations}f compares three scheduling configurations on the novice robotics task with Gemini 3.7 Flash as the primary LLM. Cross-generation and evolving overlap raise concurrency and exploration but not task performance. The former delays archive feedback, and the latter relies on an LLM judge that may pick suboptimal candidates without hardware-grounded evaluation. With Gemini 4 Argon, longer reasoning makes inference outlast simulation, so in-generation overlap alone hides 85\% and 91\% of simulation time for robotics and automotive (Fig.~\ref{fig:ablations}g). We default to in-generation overlap, which hides most simulation latency while keeping hardware-grounded feedback between generations.

\sssec{Parameter Sweeper}
The lightweight sweeper uses Flash Lite for Flash-generated solutions and Flash for Argon-generated ones, evaluating five trials with 10\% of the original time-to-live (TTL) budget. It still improves 78\% of Flash and 54\% of Argon generations (Fig~\ref{fig:ablations}h) and at least doubles scores in 3.6\%. All of this happens within the current generation, without another orchestrator--subagent cycle. Argon's stronger initial solutions likely leave less room for tuning.

\sssec{Cross generation cache.}
The first three layers reach 72.7\% reuse (Fig~\ref{fig:ablations}i). As hardware and placement stabilize and later generations mainly refine off-sensor processing (Layer 5). Frame-level caching serves 81.4\% of radar-simulation frames. Even when the overall placement changes, frames with unchanged hardware, sensor settings, and poses reuse simulated data.

\sssec{Optimizer.}
On robotic imaging, which are mostly off-sensor, before and after optimization on an NVIDIA RTX 5090 reveals Fig~\ref{fig:ablations}j where the \textit{Optimizer} LLM turns a PyTorch time-domain backprojection into a Triton kernel with an $18\times$ speedup. Across all 20 generations, optimization cuts mean execution time by 71.07\%.

%% file: conclusion.tex
\section{Conclusion}

We present \name, an agent that designs radar sensing pipelines end to end under SWaP-C constraints, representing five hardware--software layers as code with explicit dependencies and grounding its search in RF simulation, on-sensor emulation, and real edge-device execution. Across robotic, automotive, and vital-signs sensing, it approaches or exceeds human-expert designs and closed-source agents like Claude Code. More broadly, our results suggest that agents designing physical systems are limited less by writing code than by knowing what structure a working system needs and whether it will work in the world. \name's harness, tools, and benchmarks will be released to foster a research ecosystem for cyber-physical design agents, with radar as a first testbed, and to bring radar design from thousands of experts to anyone with a sensing problem.

\section{Use of AI Disclosure}
We used Anthropic Claude and OpenAI ChatGPT to assist with coding, and with organizing, editing, and proofreading the manuscript. We planned \name's design, guided and reviewed AI code and implementation. This assistance was separate from the use of AI models and agents in our system and experiments. We take full responsibility for the results reported in this manuscript.